\documentclass[12pt]{article}
\begin{document}
\begin{flushright}
CERN-TH/2002-143 \\
PNPI-2002-2484 \\
SLAC-PUB-9318 \\
hep-ph/0207297 \\
July 2002
\end{flushright}
\vspace*{1.cm}
\begin{center}
{ \large \bf High-Energy QCD  Asymptotics of \\  
   Photon--Photon Collisions}
\end{center}
\vspace*{0.3cm}
\begin{center}
{ \large Stanley~J.~Brodsky${}^{\$}$, Victor~S.~Fadin${}^{\dagger}$,
Victor~T.~Kim${}^{\& \ddagger }$, \\
 Lev~N.~Lipatov${}^{\ddagger}$
{\rm and}
Grigorii~B.~Pivovarov${}^{\S}$ } \\ 
\end{center}
\begin{center}
${}^\$ $  SLAC, Stanford, CA 94309, USA \\
${}^\dagger$  Budker Institute for Nuclear Physics,
Novosibirsk 630090, Russia \\
${}^\&$ CERN, CH-1211, Geneva 23, Switzerland \\
${}^\ddagger$ St.  Petersburg Nuclear Physics Institute,
Gatchina 188300, Russia \\
${}^\S$  Institute for Nuclear Research, Moscow 117312, Russia
\end{center}
\vspace{0.5cm}
\begin{center}
{\large \bf Abstract}
\end{center}
The high-energy behaviour of the total cross
section for highly virtual photons, as predicted by the
BFKL equation at next-to-leading order (NLO) in QCD, is discussed.
The NLO BFKL predictions, improved
by the BLM optimal scale setting, are in good agreement
with recent OPAL and L3 data at CERN LEP2.
NLO BFKL predictions for future linear colliders are presented. 
\vspace*{2.5cm}

Pis'ma ZHETF {\bf 76}, 306 (2002) [JETP Letters {\bf 76}, 249 (2002)]

\newpage

Photon--photon collisions, particularly 
$\gamma^* \gamma^*$ processes, play a special role 
in QCD~\cite{Budnev75}, since their
analysis is under much better control than the calculation of hadronic
processes, which require the input of non-perturbative hadronic
structure functions or wave functions.  In addition,
unitarization (screening) corrections due to multiple
Pomeron exchange should be less important
for the scattering of $\gamma^*$ of high
virtuality than for hadronic collisions.

The high-energy asymptotic behaviour of the $\gamma \gamma$ total cross
section in QED can be calculated~\cite{Gribov70} by an all-orders
resummation of the leading terms:
$\sigma \sim \alpha^4 s^{\omega}$, $\omega =
\frac{11}{32} \pi \alpha^2
\simeq 6 \times 10^{-5}$ 
(Fig.~1).  
However, the slowly rising
asymptotic behaviour of the QED cross section is
not apparent since large contributions come
from other sources, such as the cut of the fermion-box
contribution:
$\sigma \sim \alpha^2 (\log s)/s$ \cite{Budnev75}
(which although subleading in energy dependence, dominates the rising
contributions by powers of the QED coupling constant)
and QCD-driven processes 
(Fig.~2).
\begin{figure}[htb]
\label{fig:QED}
\vspace*{3.5cm}
\begin{center}
\includegraphics{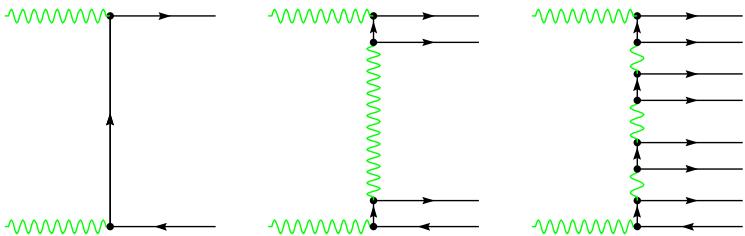}
\caption[*]{
Photon--Photon collisions in QED:
(a) electron-box diagram: $\sigma \sim \alpha^2 (\log s)/s$; 
(b) one-photon exchange diagram: $\sigma \sim \alpha^4 s^{0}$ ;
(c) a typical higher-order diagram; its resummation leads to
$\sigma \sim \alpha^4 s^{\omega}$, $\omega =
\frac{11}{32} \pi \alpha^2$ \cite{Gribov70}. 
}
\end{center}
\end{figure}

The high-energy asymptotic behaviour of hard QCD processes
is governed by the Balitsky--Fadin--Kuraev--Lipatov (BFKL) formalism
\cite{FKL,BL78}.
The highest eigenvalue, $\omega$, of the BFKL
equation \cite{FKL} is
related to the intercept of the QCD BFKL Pomeron,
which in turn governs the high-energy asymptotics
of the cross sections: $\sigma \sim
s^{\alpha_{I \negthinspace P}-1} = s^{\omega}$.
The BFKL Pomeron intercept in the leading order (LO)
turns out to be rather large:
$\alpha_{I \negthinspace P} - 1 =\omega_{LO} =
12 \, \ln2 \, ( \alpha_S/\pi )  \simeq 0.55 $ for
$\alpha_S=0.2$ \cite{FKL}.
The next-to-leading order (NLO)
corrections to the BFKL intercept have 
recently been calculated
\cite{FL}, but the results in the
$\overline{\mbox{MS}}$ scheme have a strong renormalization scale
dependence.
In Ref.~\cite{BFKLP} we used the Brodsky--Lepage--Mackenzie (BLM)
optimal scale setting procedure~\cite{BLM} to eliminate the
renormalization scale ambiguity.  (For other approaches to the NLO BFKL
predictions, see Refs.~\cite{Ciafaloni99,BFKLP} and references therein.)
The BLM optimal scale setting resums the conformal-violating
$\beta_0$-terms into the running coupling in all orders
of perturbation theory, thus preserving the conformal properties
of the theory.  The NLO BFKL predictions, as improved by 
the BLM scale setting, yields 
$\alpha_{I \negthinspace P} - 1 =\omega_{NLO} =$
0.13--0.18 \cite{BFKLP}. Strictly speaking the integral kernel of the
BFKL equation at NLO is not conformally invariant and, hence, one should
use a more accurate method for its solution (see Ref. \cite{conf}). But
in the BLM approach the dependence of the eigenvalue of the kernel
from the gluon virtuality is extremely weak \cite{BFKLP} and, therefore, 
$\omega_{NLO}$ coincides basically with the eigenvalue.  

\begin{figure}[htb]
\vspace*{3.5cm}
\begin{center}
\includegraphics{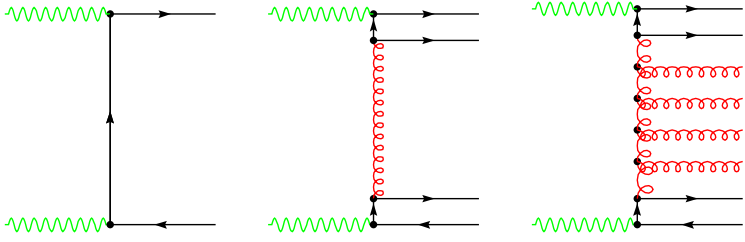}
\caption[*]{
High-energy photon-photon collisions in QCD:
(a) quark-box diagram:  $\sigma \sim \alpha^2 (\log s)/s$; 
(b) one-gluon exchange diagram: $\sigma \sim \alpha^2 \alpha_S^2 s^{0}$;
(c) a typical higher-order diagram; its resummation leads to
$\sigma \sim \alpha^2 \alpha_S^2 s^{\omega}$, $\omega_{LO} =
12 \, \ln2 \, ( \alpha_S/\pi )  \simeq 0.55 $ \cite{FKL}
and $\omega_{NLO} =$ 0.13-0.18  \cite{BFKLP}.
}
\end{center}
\label{fig:QCD}
\end{figure}
The photon--photon cross sections with LO BFKL resummation
was considered in Refs.~\cite{BL78,Bartels96,Brodsky97,Boonekamp}. 
The total cross section of two unpolarized gammas with virtualities $Q_{A}$
and $Q_B$ in the LO BFKL \cite{Brodsky97,BL78} reads as follows:
\begin{eqnarray}
\sigma(s,Q_A^2,Q_B^2) =
  \sum_{i,k = T,L}
\frac{1}{\pi \sqrt{Q_A^2 Q_B^2}} \int_{0}^{\infty}
\frac{d \nu}{2 \pi} 
 \cos \Biggl(\nu \ln \biggl(\frac{Q_A^2}{Q_B^2}\biggr)\Biggr) 
F_{i}(\nu) F_{k}(-\nu)
\Biggl( \frac{s}{s_0}\Biggr)^{\omega(Q^2,\nu)},  
\label{eqn:sigma-g}
\end{eqnarray}
with the gamma impact factors in the LO
for the transverse and longitudinal polarizations:
\begin{eqnarray}
F_T(\nu) = F_T(- \nu) = \alpha \, \alpha_S \, 
\Bigg( \sum_q e_q^2 \Bigg)  \frac{\pi}{2}
\frac{\Bigl[\frac{3}{2} - i \nu \Bigr] 
\Bigl[\frac{3}{2} + i \nu \Bigr] 
\Gamma \Big(\frac{1}{2} - i \nu \Big)^2
\Gamma \Big(\frac{1}{2} + i \nu \Big)^2}
{\Gamma (2 - i \nu) \Gamma (2 + i \nu)} \nonumber
\label{eqn:impactT}
\end{eqnarray}
\begin{eqnarray}
F_L(\nu) = F_L(- \nu) = \alpha \, \alpha_S \, 
\Bigg( \sum_q e_q^2 \Bigg) \pi
\frac{\Gamma \Big(\frac{3}{2} - i \nu \Big) 
\Gamma \Big(\frac{3}{2} + i \nu \Big) 
\Gamma \Big(\frac{1}{2} - i \nu \Big)
\Gamma \Big(\frac{1}{2} + i \nu \Big)}
{\Gamma (2 - i \nu) \Gamma (2 + i \nu)}, \nonumber
\label{eqn:impactL}
\end{eqnarray}
where a Regge scale parameter $s_0$ is proportional to
a hard scale $Q^2 \sim Q_A^2,Q_B^2$; $\Gamma$ is the Euler 
$\Gamma$-function and $e_q$ is the quark electric charge.

Although the NLO impact factor of the virtual photon is not
known \cite{IFNLO},  one can use the LO impact factor of Refs. 
\cite{Gribov70,Brodsky97},
assuming that the main energy-dependent 
NLO corrections come from the NLO
BFKL subprocess rather than from the photon impact
factors \cite{KLP,BFKLP01}.

\begin{figure}[!thb]
\vspace*{10.5cm}
\begin{center}
\includegraphics{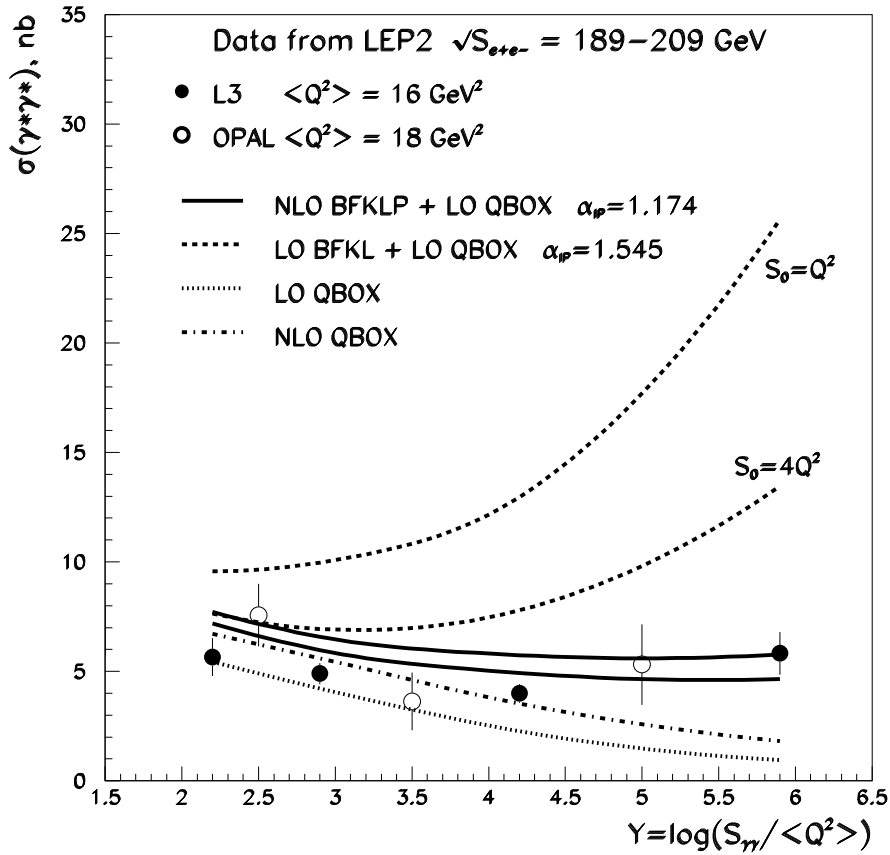}
\caption[*]{
The energy dependence of the total cross section for
highly virtual photon--photon collisions predicted
by the BLM scale-fixed NLO BFKL \cite{KLP,BFKLP01,BFKLP} 
compared with OPAL \cite{OPAL} and L3 \cite{L3} 
data from LEP2 at CERN.
The (solid) dashed curves correspond to the (N)LO BFKL predictions
for two different choices of the Regge
scale: $s_0= Q^2$  for upper curves and $s_0=4 Q^2$ for lower curves.}
\end{center}
\label{fig:CERN}
\end{figure}

Fig.~3
compares the LO and BLM scale-fixed NLO BFKL
predictions 
$\sigma \sim \alpha^2 \alpha_S^2 s^{\omega}$~\cite{BFKLP,KLP,BFKLP01}
with recent CERN LEP2 data from OPAL \cite{OPAL} 
and L3 \cite{L3}.  The spread in the curves
reflects the uncertainty in the
choice of the Regge scale parameter, which defines
the beginning of the asymptotic regime:
 $s_0=Q^2  ~ {\rm to} ~ 4Q^2 $ for LO and NLO BFKL, where $Q^2$ is the mean
virtuality of the colliding photons.  One can see from 
Fig.~3
that the agreement of the NLO BFKL predictions \cite{KLP,BFKLP01,BFKLP}
with the data is quite good.  
The sensitivity of the NLO BFKL results
to the Regge parameter $s_0$ is much smaller than in
the case of the LO BFKL.  The variation of the predictions in the value of
$s_0$ reflects uncertainties from uncalculated
subleading terms.  The parametric variation of the LO BFKL predictions
is so large that it can be, in fact, neither ruled out
nor confirmed at the energy range of CERN LEP2.

The double-logarithmic DGLAP asymptotics related 
with $\log(Q_A^2/Q_B^2)$-terms for the total photon--photon 
cross section was considered in Ref. \cite{Boonekamp}
and found to be small for the CERN LEP2 kinematical region.
The point is that most of the CERN LEP2 data \cite{OPAL,L3,ALEPH}
are collected at the approximately equal virtualities
of the colliding photons: $1/2 < Q_A^2/Q_B^2 < 2$.
It should be stressed that the soft Pomeron contribution to 
the $\gamma^\star\gamma^\star$ total cross section,
if estimated within the vector-dominance model, is proportional 
to $\sigma_{\gamma^\star\gamma^\star} 
\sim (m_V^2/Q^2)^4 \sigma_{\gamma \gamma}$ and therefore
suppressed for such highly virtual photons as those under consideration.

We also note that the NLO BFKL
predictions are
consistent \cite{BFKLP01} with data recently presented
by ALEPH \cite{ALEPH}.
In contrast, the
NLO quark-box contribution \cite{Cacciari}
underestimates the L3 data point at 
$Y\equiv\log(s_{\gamma\gamma}/ \langle Q^2 \rangle)=6$
by 4 standard deviations. Indeed, the
NLO quark-box contribution \cite{Cacciari}, calculated 
in massless approximation, can be scaled down
from general considerations with the quark masses. 
For example, at leading order, the inclusion of masses to
the quark-box diagram reduces its contribution
by 10-15$\%$ \cite{Cacciari}. 
Also, the one--gluon exchange added to the (N)LO quark-box contribution
is not sufficient to describe the data
at $Y = 6$ within (3) 4  standard deviations
(see also 
Fig.~4).

\begin{figure}[!thb]
\vspace*{12.0cm}
\begin{center}
\includegraphics{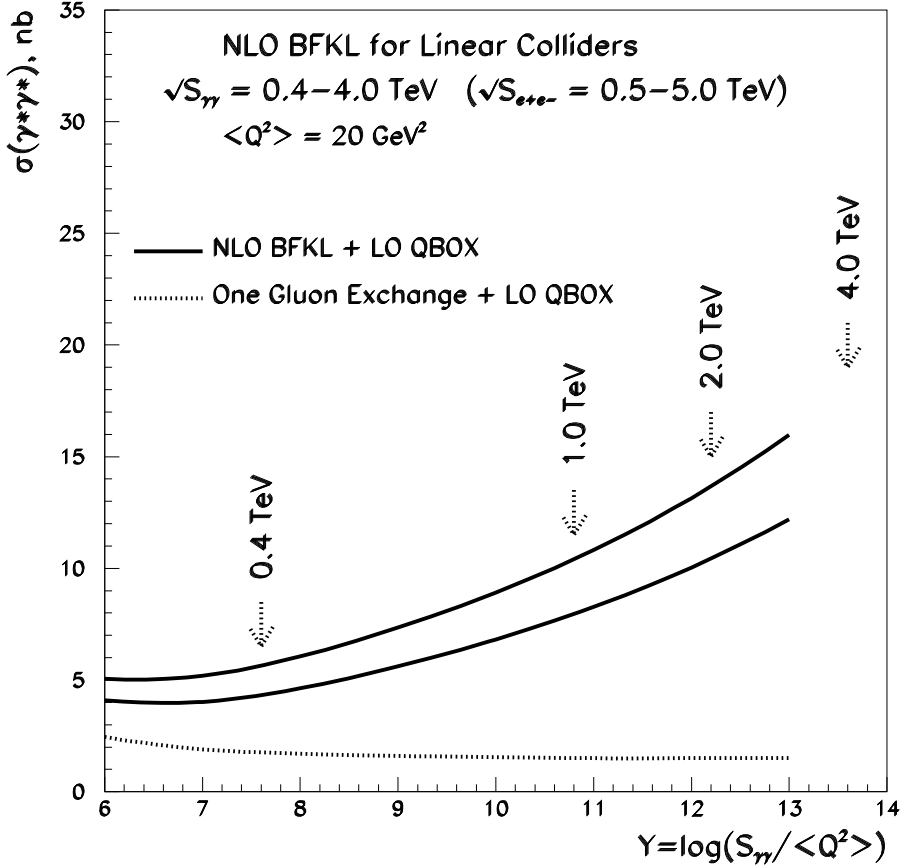}
\caption[*]{
The energy dependence of the total cross section for
virtual photon--photon collisions predicted
by the NLO BFKL  for future linear colliders.
The solid curves correspond to the BLM scale-fixed NLO BFKL predictions
with $s_0=Q^2$ (upper curve) and $s_0=4 Q^2$ (lower curve).
The dotted curve shows the one-gluon exchange contribution. }
\end{center}
\label{fig:LC}
\end{figure}

In 
Fig.~4
the BLM fixed-scale NLO BFKL 
predictions for a future linear collider 
with the photon-photon collision option
($\sqrt{s_{\gamma\gamma}} \le 0.8 \sqrt{s_{e^+e^-}}$)
under discussion \cite{LC} are shown.

The NLO BFKL phenomenology is consistent with the assumption of
small unitarization corrections in the photon--photon
scattering at large $Q^2$.  Thus one can accommodate the NLO BFKL
Pomeron intercept value 1.13--1.18 \cite{BFKLP} predicted by the BLM
optimal scale setting.  In the case of hadron scattering, 
the larger unitarization corrections~\cite{Kaidalov86}
lead to a smaller effective Pomeron intercept value, 
about 1.10~\cite{Cudell00}.

In summary, highly virtual photon--photon collisions
provide a very unique opportunity to test high-energy asymptotics
of QCD.  The NLO BFKL predictions for the $\gamma^* \gamma^*$ 
total cross section, with the renormalization scale fixed
by the BLM procedure, show good agreement with the recent data from
OPAL \cite{OPAL} and L3 \cite{L3} at CERN LEP2.
The obtained results can be very important for  
future lepton and photon colliders.

The authors thank V.~P.~Andreev, A.~Bohrer, A.~De~Roeck, J.~R.~Ellis, 
J.~H.~Field, I.~M.~Ginzburg, A.~B.~Kaidalov, V.~A.~Khoze, 
M.~Kienzle-Focacci, M.~Krawczyk, C.-N.~Lin,
V.~A.~Schegelsky, V.~G.~Serbo, M.~Przybycie\'n, A.~A.~Vo\-ro\-byov 
and M.~Wadhwa for helpful discussions.
This work was supported in part by the Russian
Foundation for Basic Research (RFBR), the INTAS Foundation, 
the U.S. National Science Foundation, and the U.S. Dept. of Energy
under contract No. DE-AC03-76SF00515.

\end{document}